\documentclass[debug,overfull]{epl}

\usepackage{graphics,epsf,psfig}
\usepackage{amssymb,amsfonts,amsmath}
\usepackage{bbm}

\def\ket#1{\mathinner{|{#1}\rangle}}
\def\rbra#1{\mathinner{\langle{#1}\parallel}}
\def\rket#1{\mathinner{\parallel{#1}\rangle}}

\def\Bra#1{\left<#1\right|}
\def\Ket#1{\left|#1\right>}
\newcommand{\Hhf}{\ensuremath{\mathrm{H}_{\mbox{\tiny hf}}}}
\newcommand{\HSt}{\ensuremath{\mathrm{H}_{\mbox{\tiny St}}}}

\newcommand{\E}{\mathbb{E}}

\newcommand{\Hz}{\mbox{Hz}}

\newcommand{\cm}{\mbox{cm}}
\newcommand{\kV}{\mbox{kV}}

\title{Reconciliation of experimental and theoretical electric \\ tensor
polarizabilities of the cesium ground state} \shorttitle{Tensor polarizabilities in cesium}
\author{S.~Ulzega \thanks{email: \email{simone.ulzega@unifr.ch}}
   \and A.~Hofer \and P.~Moroshkin \and A.~Weis}
\shortauthor{S.~Ulzega \etal}

\institute{Physics Department, University of Fribourg, \\ Chemin
du Mus\'ee 3, 1700 Fribourg, Switzerland }


\begin{document}

\maketitle

\begin{abstract}
We present a new theoretical analysis of the strongly suppressed
$F$- and $M$-dependent Stark shifts of the Cs ground state
hyperfine structure. Our treatment uses third order perturbation
theory including off-diagonal hyperfine interactions not
considered in earlier treatments. A numerical evaluation of the
perturbation sum using bound states up to n=200 yields ground
state tensor polarizabilities $\alpha_2(6S_{1/2},F)$ which are in
good agreement with experimental values, thereby bridging the
40-year-old gap between experiments and theory. We have further
found that the tensor polarizabilities of the two ground state
hyperfine manifolds have opposite signs, in disagreement with an
earlier derivation. This sign error has a direct implication for
the precise evaluation of the blackbody radiation shift in primary
frequency standards.
\end{abstract}

\section{Introduction}
Since its discovery, the Stark effect, i.e., the interaction of an
atom with an external electric field $\mathbb{E}$, has played an
important role in the spectroscopic investigation of atomic
structure.
Because of parity conservation, the Stark effect arises only in
second order perturbation theory for atoms with non-degenerate
orbital angular momentum states.
The energy shift $\Delta E(\gamma)$ of a magnetic sublevel
$\ket{\gamma}=\ket{nL_J, F,M}$ of the hyperfine structure is
conventionally parametrized in terms of a polarizability
$\alpha(\gamma)$ as
\begin{equation}\label{eq:alphaDef}
\Delta E(\gamma) = -\frac{1}{2}\,\alpha(\gamma)\,\mathbb{E}^2\,.
\end{equation}
In second order perturbation theory the polarizability can be
decomposed~\cite{angel_sandars} into a scalar polarizability,
$\alpha^{(2)}_{0}$, which leads to an $F$- and $M$-independent
level shift, and a tensor polarizability, $\alpha^{(2)}_{2}$,
which yields $F$- and $M$-dependent energy shifts.
Because of rotational symmetry, the tensor polarizability vanishes
for the spherical $nS_{1/2}$ and $nP_{1/2}$ states in the alkalis.
As a consequence, the second order Stark effect in the alkali
ground state does not affect its hyperfine splitting, nor does it
lift the Zeeman degeneracy of a given hyperfine level.
However, Haun and Zacharias observed  already in
1957~\cite{haun_zacharias} that a static electric field does
induce a quadratic Stark shift of the hyperfine transition
frequency ($F$-dependent Stark effect), and in 1964, Lipworth and
Sandars~\cite{lipworth_sandars} demonstrated that a static
electric field also lifts the Zeeman degeneracy within the $F=4$
sublevel manifold of the cesium ground state (an
$\left|M\right|$-dependent effect).
Improved measurements were performed later by Carrico
\etal~\cite{carrico_tenspol} and Gould
\etal~\cite{gould_tenspol}, and were recently confirmed by our
own measurements~\cite{ospelkaus_tenspol,ulzega,ulzegaPhDthesis}.
%
%
%$\alpha_{0}^{(2)}=99.78(15)\:\kHz/(\kV/\cm)^{2}$~\cite{amini_scalpol}.
%
In 1967, Sandars~\cite{sandars} showed that the $F$- and
$\left|M\right|$-dependence of the Stark effect can be explained
when the perturbation theory is extended to third order after
including the hyperfine interaction.
His theoretical expressions for the third order tensor
polarizabilties $\alpha_{2}^{(3)}$ were evaluated numerically
in~\cite{lipworth_sandars} and~\cite{gould_tenspol} under
simplifying assumptions.
This yielded (absolute) values which were systematically larger
than the experimental values for the five investigated alkali
isotopes~\cite{gould_tenspol}.
It is worth noting that in cesium the level shifts due to
$\alpha_{2}^{(3)}$ are almost seven orders of magnitude smaller
than the common shift of the ground state levels due to the second
order scalar polarizability $\alpha_{0}^{(2)}$.
While the scalar Stark shift is understood at a level of
$\left(1\mbox{--} 2
\right)\cdot10^{-3}$~\cite{amini_scalpol,derevianko_scalpol,zhou_scalpol},
there has so far been no satisfactory theoretical description of
the tensor polarizabilities, i.e., of the
$\left|M\right|$-dependent alterations of the scalar effect.

In this paper we extend the previous theoretical treatment of the
third order effects by including both diagonal and off-diagonal
hyperfine interactions in the perturbation expansion.
The numerical evaluation of the perturbation sum then yields a value
of the Cs ground state tensor polarizability $\alpha_2(6S_{1/2},F=4)$
which is in very good agreement with the experimental
results~\cite{carrico_tenspol, gould_tenspol, ospelkaus_tenspol,
ulzega, ulzegaPhDthesis}, thereby bridging the 40-year-old
gap~\cite{ulzegaPhDthesis} between experiments and theory.
We have also identified an error in Sandars'
results~\cite{sandars} concerning the relative sign of the tensor
polarizabilities of the two ground state hyperfine levels.
We discuss the relevance of this sign error for the determination
of the black-body radiation shift of primary frequency standards
from measurements of the static Stark shift of the Cs clock
frequency.

\section{Theoretical model}

As shown by Sandars~\cite{sandars}, the $F$- and $M$-dependent
Stark shifts can be explained by a third order perturbation
treatment, in which the Stark interaction,
$\HSt = - \overrightarrow{d}\cdot\overrightarrow{\E}$,
and the dipole-dipole ($\Hhf^{\rm{d}}$) and electric quadrupole
($\Hhf^{\rm{q}}$) hyperfine interactions,
are treated on an equal footing.  The contributions of $\HSt$ and
$\Hhf = \Hhf^{\rm{d}}+\Hhf^{\rm{q}}$
can be expressed in terms of irreducible tensor operators
\begin{align}
\HSt & =\left|e\right|r\mathbf{C}^{(1)}\cdot\mathbf{E}^{(1)}
\label{eq:HSt}\\
\Hhf^{\rm{d}} & =
      \left( a_{l,j} \left\{ \mathbf{L}^{(1)}-\sqrt{10}
                          \bigl[ \mathbf{C}^{(2)}\times
                                 \mathbf{S}^{(1)}
                          \bigr]^{(1)}
                    \right\} + a_{s} \mathbf{S}^{(1)}
     \right) \cdot \mathbf{I}^{(1)}\:,
\label{eq:Hhf1}\\
\Hhf^{\rm{q}} & = b_{Q}
\mathbf{C}^{(2)}\cdot
               \bigl[\mathbf{I}^{(1)}\times\mathbf{I}^{(1)}
               \bigr]^{(2)}\:,
\label{eq:Hhf2}
\end{align}
where $\mathbf{L}$, $\mathbf{S}$, and $\mathbf{I}$ are
% rank~1 tensor operators
vector operators associated with the orbital angular momentum, the
electronic spin, and the nuclear spin, respectively, and where the
$\mathbf{C}^{(k)}$ are normalized spherical harmonics of rank~k.
The first and second terms of Eq.~(\ref{eq:Hhf1}) represent the
magnetic interaction of the nuclear magnetic moment with the
orbital and electron spin dipole moments respectively
(dipole-dipole interactions), while the third term represents the
Fermi contact interaction and Eq.~(\ref{eq:Hhf2}) is the electric
quadrupole interaction.
The corresponding coupling constants are $a_{l,j}$, $a_{s}$, and $b_{Q}$.
The contact interaction term has non-zero matrix elements between
S-states ($L=0$) only, while the first two terms of
Eq.~(\ref{eq:Hhf1}) apply to states with $L>0$. The electric
quadrupole term of Eq.~(\ref{eq:Hhf2}) requires $L > 1$ and $J >
1$, and for Cs $nP_{3/2}$ states its matrix elements are two
orders of magnitude smaller than all other contributions.

The third order energy perturbation of the sublevel
$\ket{\alpha} = \ket{6S_{1/2},F,M}$ is given by
\begin{equation}
\Delta E^{(3)}(\alpha) =\sum_{\beta\neq\alpha, \gamma\neq\alpha}
\frac{\Bra{\alpha} W \Ket{ \beta}
      \Bra{ \beta} W \Ket{\gamma}
      \Bra{\gamma} W \Ket{\alpha}}
     {(E_{\alpha}-E_{\beta})(E_{\alpha}-E_{\gamma})}
- \Bra{\alpha}W \Ket{\alpha} \sum_{\beta \neq \alpha} \frac{\left|
\Bra{\beta}W\Ket{\alpha} \right|^{2}}{(E_{\alpha}-E_{\beta})^2} \:,
\label{eq:thirdordergeneral}
\end{equation}
where $E_{\beta}$ and $E_{\gamma}$ are the unperturbed state energies.
Of all the terms obtained by substituting $W=\HSt+\Hhf$  into
Eq.~(\ref{eq:thirdordergeneral}) only those proportional to $\E^2$
give nonzero contributions to the Stark interaction, and the sums
have to be carried out over all states according to the selection
rules imposed by the operators.

We first address the second term of
Eq.~(\ref{eq:thirdordergeneral}), whose only non-vanishing
contribution is
\begin{equation}
- \Bra{\alpha}\Hhf \Ket{\alpha} \sum_{\beta \neq \alpha}
\frac{\left| \Bra{\beta}\HSt\Ket{\alpha}\right|^{2}}
     {(E_{\alpha}-E_{\beta})^2} \equiv - E_{hf}(\alpha) \sum_{\beta \neq \alpha}
\frac{\left| \Bra{\beta}\HSt\Ket{\alpha}\right|^{2}}
     {(E_{\alpha}-E_{\beta})^2} \:,
\label{eq:thirdorderscalterm}
\end{equation}
where $\ket{\beta}=\ket{nP_{j},f,m}$.  Diagram $\textbf{A}$ of
Fig.~\ref{fig:fig1} shows a graphical representation of this term
in which the Fermi ground-state contact interaction appears as a
multiplicative factor, making this contribution $F$-dependent.
Comparing the sum in Eq.~(\ref{eq:thirdorderscalterm}) to the
expression
\begin{equation}
\Delta E^{(2)}(\alpha) =\sum_{\beta\neq\alpha} \frac{\left|
\Bra{\beta} \HSt \Ket{ \alpha}\right|^{2}}
     {E_{\alpha}-E_{\beta}} \:,
\label{eq:secondordergeneral}
\end{equation}
for the second order scalar Stark effect, one sees that the
F-dependent third order shift is suppressed by a factor on the
order of
$E_{hf}(6S)/(E_{6P}-E_{6S})\approx 10^{-5}$.
Following the definition of Eq.~(\ref{eq:alphaDef}) the
contribution of Eq.~(\ref{eq:thirdorderscalterm}) can be
parametrized in terms of an $F$-dependent third order scalar
polarizability $\alpha_{0}^{(3)}(6S_{1/2},F)$. This $F$-dependent
term gives the main contribution to the Stark shift of the
hyperfine transition frequency.

We next address the first term of
Eq.~(\ref{eq:thirdordergeneral}). The nonvanishing contributions
involving \emph{diagonal} matrix elements of $\Hhf$ are given by
\begin{equation}
\sum_{\beta\neq\alpha}\Bra{ \beta} \Hhf \Ket{\beta}
\frac{\left|\Bra{\beta}\HSt\Ket{\alpha} \right|^{2}}
     {(E_{\alpha}-E_{\beta})^{2}} \equiv \sum_{\beta\neq\alpha} E_{hf}(\beta)
\frac{\left|\Bra{\beta}\HSt\Ket{\alpha} \right|^{2}}
     {(E_{\alpha}-E_{\beta})^{2}} \:,
\label{eq:thirdordertensterm}
\end{equation}
where $\ket{\beta}=\ket{nP_{j},f,m}$. They are represented by
diagram $\textbf{B}$ of Fig.~\ref{fig:fig1} and are suppressed by
a factor on the order of
$E_{hf}(6P)/(E_{6P}-E_{6S})\approx 10^{-7}$
with respect to the second order scalar shifts. The electric field
dependent factor of Eq.~(\ref{eq:thirdordertensterm}) has only a
rank $k=0$ (scalar) contribution, while the dipole-dipole
[Eq.~(\ref{eq:Hhf1})] and the electric quadrupole
[Eq.~(\ref{eq:Hhf2})] parts of the $\Hhf$ factor in
Eq.~(\ref{eq:thirdordertensterm}) have the rotational symmetries
of $k=0,2$ and $k=2$ tensors, respectively.
The contraction of the hyperfine and the Stark interactions in
Eq.~(\ref{eq:thirdordertensterm}) thus yields both scalar and
second rank tensor contributions.
The scalar part of Eq.~(\ref{eq:thirdordertensterm}) has the same
$F$-dependence~\cite{ulzegaPhDthesis} as $\alpha_{0}^{(3)}(F)$,
and gives a correction to the latter on the order of 1\%, while
the second rank tensor part has an $F$- and $M$-dependence, which
can be parametrized in terms of a third order tensor
polarizability $\alpha_{2}^{(3)}(6S_{1/2},F)$.

The total third order polarizability can thus be written as
\begin{equation}
\alpha^{(3)}(6S_{1/2},F,M)=\alpha_{0}^{(3)}(6S_{1/2},F) + \alpha_{2}^{(3)}(6S_{1/2},F) \frac{3M^{2}\!-\!F(F\!+\!1)}{2 I(2I+1)} f(\theta) \,,
\label{eq:globalpol}
\end{equation}
where the dependence on the angle $\theta$ between the electric
field and the quantization axis is given by
$f(\theta)=3\cos^{2}\theta-1$.
Equations~(\ref{eq:thirdorderscalterm},\ref{eq:thirdordertensterm})
can be reduced by applying the Wigner-Eckart theorem and angular
momentum decoupling rules.
For cesium ($I=7/2$) the explicit $F$- and $M$-dependences of the
third order polarizabilities for  $\theta=0$ are
\begin{subequations}
\label{eq:alpha2Sandras}
\begin{eqnarray}
\alpha^{(3)}(F\!=\!4,M) & = & a_0+\left(a_1+a_2\right)
\frac{3M^{2}-20}{28}\:,
\label{subeq:alpha2F=4}\\
\alpha^{(3)}(F\!=\!3,M) & = & -\frac{9}{7}\,a_0+\left(-a_1+
\frac{5}{3}\,a_2\right) \frac{3M^{2}-12}{28}\:.
\label{subeq:alpha2F=3}
\end{eqnarray}
\end{subequations}

The scalar part can be expressed in terms of radial integrals
\begin{equation}
a_{0}= \frac{7}{54} \sum_{n}
       \left[ \mathcal{C}_{nP_{1/2}}
              \left( 3A_{6S_{1/2}} +   A_{nP_{1/2}} \right)
              + 2 \mathcal{C}_{nP_{3/2}}
              \left( 3A_{6S_{1/2}} - 5 A_{nP_{3/2}} \right)
       \right]
\:,
\label{a0}
\end{equation}
where $A_{6S_{1/2}}$ and $A_{nP_{j}}$ are the hyperfine coupling constants and where
\begin{equation}
\mathcal{C}_{nP_{j}}=\frac{e^{2}\left| R_{\mathrm{6S,nP_{j}}}
\right|^{2}}{\left(E_{6S_{1/2}}-E_{nP_{j}}\right)^{2}}\: .
\end{equation}
$R_{\mathrm{6S,nP_{j}}}$ is the radial integral between the ground
state and the excited $\Ket{nP_{j}}$ state.
The Fermi-contact interaction (proportional to $A_{6S_{1/2}}$)
provides the dominant contribution to $a_0$ which, as mentioned,
also has a small contribution (proportional to $A_{nP_{j}}$) from
the scalar part of the magnetic dipole interactions.
Eqs.~(\ref{eq:thirdorderscalterm}) and (\ref{eq:thirdordertensterm}) can be expressed in a similar way as
\begin{equation}
a_{1} = - \frac{7}{54} \sum_{n}
       \left[
          \left( 2A_{nP_{1/2}} \mathcal{C}_{nP_{1/2}}
               - 5A_{nP_{3/2}} \mathcal{C}_{nP_{3/2}} \right)
        + \left( 2A_{nP_{1/2}} \mathcal{C}_{nP_{1/2}}
               +  A_{nP_{3/2}} \mathcal{C}_{nP_{3/2}} \right)
       \right] \:,
\label{eq:a1}
\end{equation}
where we have explicitely separated the contribution of the
orbital part of the interaction (first term) from that of the spin
dipolar part (second term), and
\begin{equation}
a_{2}= \frac{1}{9} \sum_{n} B_{nP_{3/2}}\mathcal{C}_{nP_{3/2}} \:,
\label{eq:a2}
\end{equation}
where $B_{nP_{3/2}}$ is the quadrupole hyperfine coupling
constant.

Equations~(\ref{eq:alpha2Sandras}) closely resemble the
expressions derived by Sandars~\cite{sandars}, except for the
negative sign of the $a_1$ term in Eq.~(\ref{subeq:alpha2F=3})
which is positive in Sandars' work.
To our knowledge, this sign cannot be derived from any prior
experiment.
We have recently confirmed experimentally~\cite{ulzega} that the
tensor polarizabilities of the $F=3$ and $F=4$ states have indeed
opposite signs as predicted by our calculation.
This sign error has remained unnoticed in the literature for
almost 40 years and we will come back to its relevance for atomic
clocks below.

\section{Numerical evaluation of the tensor polarizability}
\label{subsect:Theory:NumEv}

The tensor polarizability ($a_1$ term) of the $F\!=\!4$ state of
cesium was evaluated in~\cite{gould_tenspol} by considering only
diagonal matrix elements of $\Hhf$ for the $6S_{1/2}$ and the
$6P_{J}$ states. The contribution of the orbital magnetic dipole
hyperfine interaction [first term in Eq.~(\ref{eq:a1})] was
neglected, as well as the spin-orbit splitting in the denominators
of Eq.~(\ref{eq:thirdordergeneral}).  The authors assumed
furthermore that $A_{6P_{1/2}}/A_{6P_{3/2}} = 5$, valid for
one-electron atoms, while for Cs the corresponding ratio of
experimental values is~5.8.  Under the latter two approximations
one can factor the second order scalar polarizability
$\alpha_{0}^{(2)}$ out of the expression for the tensor
polarizability.
Those simplifying assumptions yielded the $\alpha_{2}^{(3)}(F=4)$
value represented as point (f) in Fig.~\ref{fig:fig2}, in
disagreement with the experimental results.
We have reevaluated the result by including the orbital part of
the hyperfine interaction and then rescaling the value of
~\cite{gould_tenspol} by using the most recent experimental value
of $\alpha_{0}^{(2)}$~\cite{amini_scalpol}.  This yields the value
(f') in Fig.~\ref{fig:fig2}, thus increasing the gap between
theory and experiments.  In a second, more precise, calculation we
dropped all the simplifying assumptions mentioned
above~\cite{ulzegaPhDthesis}, still keeping diagonal matrix
elements only, and numerically evaluated
Eqs.~(\ref{eq:a1},\ref{eq:a2}) by using recent experimental
values~\cite{rafac_rme} of the reduced dipole matrix elements
$\rbra{6S_{1/2}}d \rket{6P_{j}}$.
As a result, the discrepancy with experiments becomes even larger
[point (f'') in Fig.~\ref{fig:fig2}], and does not change
significantly when extending the perturbation sum to $nP_J$ states
with $n>6$.

\section{Off-diagonal hyperfine interaction}
\label{sect:OffDiag}

\begin{figure}[t]
\onefigure[scale=0.55]{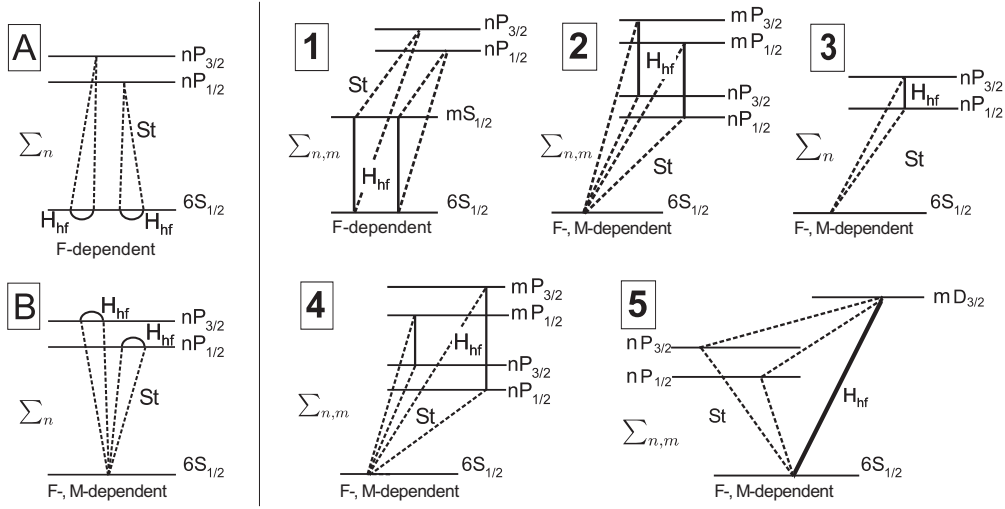}
\caption{Contributions of diagonal (\textbf{A} and \textbf{B}) and
off-diagonal (\textbf{1}\mbox{--}\textbf{5}) hyperfine matrix
elements to the third order Stark effect. Dotted/solid lines
represent the Stark and hyperfine interactions respectively. The
relative contributions of the diagrams are given in the text.}
\label{fig:fig1}
\end{figure}

All calculations described above considered only diagonal
hyperfine matrix elements.
However, the first term in Eq.~(\ref{eq:thirdordergeneral}) allows
also off-diagonal hyperfine terms.
Figure~\ref{fig:fig1} gives a schematic overview of all possible
off-diagonal configurations compatible with the hyperfine and
Stark operator selection rules.
It is interesting to note that some off-diagonal hyperfine matrix
elements (diagrams \textbf{1} and \textbf{2}, Fig.~\ref{fig:fig1})
were already considered by Feichtner \etal~\cite{feichtner} in
their calculation of the clock transition Stark shift, but for
unknown reasons such terms were never considered in the tensor
polarizability calculation.

The contact interaction selection rule $\Delta L = 0$ admits only
off-diagonal matrix elements between the ground state and higher
lying $S_{1/2}$ states (diagram \textbf{1}, Fig.~\ref{fig:fig1}).
\begin{figure}[b]
\onefigure[scale=0.85]{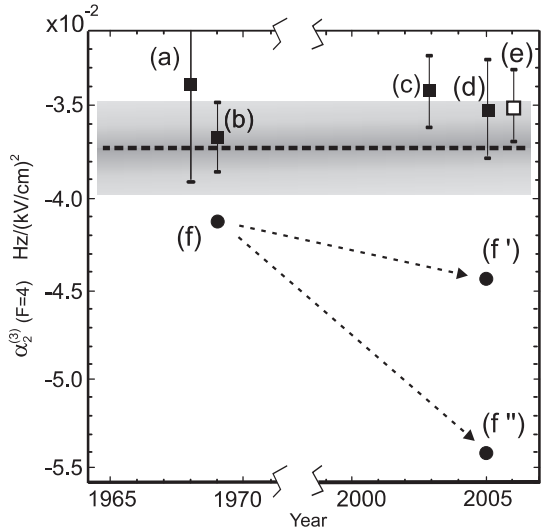} \caption{The
third order tensor polarizability of the $F\!=\!4$ Cs ground
state.  The filled squares ({\tiny$\blacksquare$}) represent
experimental values of Carrico
\etal~\protect\cite{carrico_tenspol}(a), Gould
\etal~\protect\cite{gould_tenspol}(b), Ospelkaus
\etal~\protect\cite{ospelkaus_tenspol}(c) and Ulzega
\etal~\protect\cite{ulzega}(d).  The empty square
({\tiny$\square$}) (e) represents their weighted average.  The
dots (\textbullet) represent the theoretical value
from~\protect\cite{gould_tenspol}(f), and our reevaluations (f',
f'') discussed in the text.  The dotted horizontal line is the
result of the present work with its uncertainty (shaded band).}
\label{fig:fig2}
\end{figure}
The orbital and spin dipolar magnetic interactions also obey
$\Delta L = 0$ and have nonvanishing off-diagonal matrix elements
of the form $\Bra{\beta_{1}} \Hhf \Ket{\beta_{2}}$ where
$\Ket{\beta_{i}}=\Ket{n_{i}P_{j_{i}}}$ (diagrams \textbf{2},
\textbf{3} and \textbf{4}).
Due to the second rank tensor character of $\mathbf{C}^{(2)}$, the
spin dipolar term also couples states with $\Delta L = \pm 2$ and
can thus contribute to the third order Stark effect with
off-diagonal matrix elements between the ground state and
$D_{3/2}$ states (diagram \textbf{5}).
We have used the Schr\"odinger equation with a statistical
Thomas-Fermi model potential to calculate the relevant wave functions
of the free Cs atom.
Corrections for the dipolar and quadrupolar core polarization as well
as spin-orbit interaction with a relativistic correction factor
following~\cite{norcross} were included.
The electric dipole and hyperfine matrix elements were calculated
using the Schr\"odinger wave functions for all matrix elements for
which no precise values could be found in the literature.
In this way we evaluated all diagrams in Fig.~\ref{fig:fig1} by
running the summation indices $m$ and $n$ up to 200.

\section{Relative contributions and numerical results}
\label{sect:OffDiag:NumRes}

Diagrams $\textbf{A}$ and $\textbf{1}$ of Fig.~\ref{fig:fig1}
yield only $F$-dependent energy shifts, and thus do not contribute
to the tensor polarizability. All other diagrams produce $F$- and
$M$-dependent effects.
The relative importance with which diagrams $\textbf{B}$ and
$\textbf{2}\mbox{--} \textbf{5}$ contribute to $\alpha_{2}^{(3)}$
is +145, +99, -40, -3, and -101\%, respectively.
A~numerical evaluation of the perturbation sum with all
diagonal and off-diagonal matrix elements mentioned above gives
\begin{equation}
\alpha_{2}^{(3)}(F=4)=-3.72(25)\times
10^{-2}\:\Hz/(\kV/\cm)^{2}\:, \label{tenspolfinalvalue}
\end{equation}
for the tensor polarizability, in which the contribution of $a_2$
is $2\times 10^{-5}\:\Hz/(\kV/\cm)^{2}$.
This result is shown as a dotted line in Fig.~\ref{fig:fig2}
together with previous theoretical and experimental results.
We estimate the uncertainty of our calculated value to be 7\%,
based on the precision ($2\mbox{--}8 \%$) with which our
Schr\"odinger solutions can reproduce measured dipole matrix
elements and hyperfine splittings, and considering that some (more
precise) experimental values were included in the calculations,
whenever they were available. We have also verified by explicit
calculations using continuum wave functions that continuum states
contribute only at a level of $10^{-4}$ to the diagrams relevant
for $\alpha_2^{(3)}$ ($\sim a_1$), while it was recently
shown~\cite{angstmann,beloy} that the continuum contributes
significantly ($\approx 10\%$) to diagram $\textbf{1}$, which
dominates the BBR shift via the $\alpha_0^{(3)}$ ($\sim a_0$) term
in Eq.~(\ref{eq:newclockshift})~\cite{ulzega_arXiv}.

The corrected sign of the $a_1$ terms has an important implication
for frequency standards.  From Eqs.~(\ref{eq:alpha2Sandras}) the
static Stark shift of the Cs clock frequency is given by
\begin{equation}
\Delta\nu_{00} = \Delta\nu\left(6S_{1/2},4,0\right)
                -\Delta\nu\left(6S_{1/2},3,0\right)
               = -\frac{1}{2}\left[ \frac{16}{7}\,a_{0}
                                   -\frac{4}{7}\,a_{1}\,f\left(\theta\right)
                            \right] \mathbb{E}^2  \:.
\label{eq:newclockshift}
\end{equation}
One of the leading systematic shifts of the clock transition
frequency is due to the interaction of the atoms with the
\emph{dynamic} Stark interaction induced by the blackbody
radiation (BBR) field.
This shift can be calculated from Eq.~(\ref{eq:newclockshift}) by
averaging $\mathbb{E}^2$ over the Planck
spectrum~\cite{angstmann,beloy,itano}.
Because of the isotropy of the blackbody radiation the $\theta$
dependence in Eq.~(\ref{eq:newclockshift}) vanishes, so that the
BBR shift is determined by the scalar part $a_{0}$ only.
The BBR shift coefficient can be deduced from the measured shift
of the clock transition frequency induced by a \emph{static}
electric field. For this one has to know the value of the $a_1$
term in Eq.~(\ref{eq:newclockshift}) which is affected by the sign
error, the coefficient of $a_1$ being $-\frac{1}{7}$ when derived
from Sandars' formula and $-\frac{4}{7}$ from our calculation.
The most precise measurement of the static Stark shift was
performed by Simon \etal~\cite{simon} from which the authors
extracted $-\frac{8}{7} a_{0}=-2.273(4)\: kHz/(kV/cm)^2$, while
our evaluation with the correct sign yields $-\frac{8}{7}
a_{0}=-2.281(4) \: kHz/(kV/cm)^2$.
The correction of the sign error thus changes the BBR shift rate
by a value which is twice as large as the reported experimental
uncertainty.

\section{Conclusions}
\label{sect:conclusions}

We have extended a previous treatment of the Stark interaction by
including off-diagonal hyperfine matrix elements in the third
order perturbation expansion.  Our calculation of the tensor
polarizability yields a good agreement with all experimental data,
thereby removing a 40-year-old discrepancy.  A sign error
identified in Sandars' model leads to a new expression for the
static Stark shift of the hyperfine transition frequency which
requires Cs primary frequency standards to be corrected at a level
of $6 \times 10^{-17}$, which is below their present accuracy.

\acknowledgments We acknowledge funding by the Swiss National
Science Foundation (grant 200020--103864) and useful discussions
with V.~V.~Flambaum, A.~Derevianko and M.-A.~Bouchiat.

\end{document}